\documentclass[prd,preprint,nofootinbib]{revtex4}
\usepackage{graphicx}
\usepackage{amssymb,amsmath}

\begin{document}
\newcommand{\gsim}{ \mathop{}_{\textstyle \sim}^{\textstyle >} }
\newcommand{\lsim}{ \mathop{}_{\textstyle \sim}^{\textstyle <} }
\newcommand{\vev}[1]{ \left\langle {#1} \right\rangle }
\newcommand{\del}{\partial}
\newcommand{\bear}{\begin{array}}  \newcommand{\eear}{\end{array}}
\newcommand{\bea}{\begin{eqnarray}}  \newcommand{\eea}{\end{eqnarray}}
\newcommand{\beq}{\begin{equation}}  \newcommand{\eeq}{\end{equation}}
\newcommand{\bef}{\begin{figure}}  \newcommand{\eef}{\end{figure}}
\newcommand{\bec}{\begin{center}}  \newcommand{\eec}{\end{center}}
\newcommand{\non}{\nonumber}  \newcommand{\eqn}[1]{\beq {#1}\eeq}
\newcommand{\la}{\left\langle} \newcommand{\ra}{\right\rangle}
\def\lrf#1#2{ \left(\frac{#1}{#2}\right)}
\def\lrfp#1#2#3{ \left(\frac{#1}{#2}\right)^{#3}}

\begin{flushright}
\hfill hep-ph/yymmnnn\\
\hfill DESY 07-020\\
\end{flushright}

\title{
Increasing the effective number of neutrinos with decaying particles
}

\author{Kazuhide Ichikawa, Masahiro Kawasaki, Kazunori Nakayama, Masato Senami}
\affiliation{
Institute for Cosmic Ray Research, University of Tokyo,
Kashiwa 277 8582, Japan}
\author{Fuminobu Takahashi}
\affiliation{Deutsches Elektronen Synchrotron DESY, Notkestrasse 85,
22607 Hamburg, Germany}

\date{\today}

\vskip10mm
\begin{abstract}
We present models of decaying particles for increasing the effective
number of neutrinos $N_\nu$ after big bang nucleosynthesis but before
the structure formation begins. We point out that our scenario not only solves
the discrepancy between the constraints on $N_\nu$ from these two epochs,
but also provides a possible answer to deeper inconsistency in the
estimation of the matter power spectrum amplitude at small scales,
represented by $\sigma_8$, between the WMAP and some small scale
matter power measurements such as the Lyman-$\alpha$ forest and weak
lensing.  We consider (a) saxion decay into two axions; (b) gravitino
decay into axino and axion; (c) Dirac right-handed sneutrino decay
into gravitino and right-handed neutrino.
\end{abstract}
\maketitle

\section{Introduction}
\label{sec:1}
Observations of the cosmic microwave background (CMB), galaxy clustering, 
weak gravitational lensing and Lyman-$\alpha$ forest and so on strongly 
support the cosmological structures being formed in a universe described 
by the power-law $\Lambda$CDM model. Specifically, recent advancement in
those observations enabled us to measure the matter power spectrum from the
horizon scale down to about 1\,Mpc in a very precise manner.

However, as the data accumulate owing to the recent observations for
instance by the WMAP
\cite{Spergel:2006hy,Page:2006hz,Hinshaw:2006ia,Jarosik:2006ib} and
SDSS
\cite{Tegmark:2003uf,Tegmark:2006az,McDonald:2004eu,McDonald:2004xn},
possible tensions among different data sets are indicated. This is
most easily seen in the term $\sigma_8$, the normalization of the
matter power spectrum at 8$h^{-1}$\,Mpc, where $h$ is the Hubble
parameter. Namely, the value of $\sigma_8$ derived from the WMAP
three-year data is slightly lower than that derived from the latest
analyses of the Lyman-$\alpha$ forest \cite{Viel:2006yh,Seljak:2006bg},
weak lensing \cite{Hoekstra:2005cs,Massey:2007gh} and strong lensing
(``giant arc") statistics \cite{Li:2006tk}. The discrepancy surfaced
when the WMAP data were updated from the first year data to the three-year
data, with significant decrease in the best fit value of $\sigma_8$
(decreased from $0.92 \pm 0.10$ to $0.761^{+0.049}_{-0.048}$)
\cite{Spergel:2003cb,Spergel:2006hy}. 

It is true that all these measurements which favor higher $\sigma_8$
than the WMAP3 value are likely to suffer more from systematic errors
than the WMAP experiments, but when the ongoing efforts can succeed
in decreasing the systematics, they would be more suitable for measuring
$\sigma_8$ than the CMB experiments. Therefore, we can expect that we
will obtain sufficient information to know whether the tension is
solved by some systematics not yet accounted for or we have to invoke
non-standard cosmology. We, in this paper, assume that the latter case
is true and the present discrepancy between the WMAP3 and the
observations at smaller scales is real. 

Then, what kind of non-standard ingredient do we need ? Actually, this
has been already hinted at in the Lyman-$\alpha$ forest analysis of
Ref.~\cite{Seljak:2006bg}. In Ref.~\cite{Seljak:2006bg}, extensive
cosmological parameter estimation was conducted using the latest data
set consists of the CMB, galaxy clustering and the Lyman-$\alpha$
forest.  They tested a wide range of cosmological models other than the
flat $\Lambda$CDM model with the adiabatic power-law primordial power
spectrum by placing constraints on the tensor mode, the running of the
spectral index, massive neutrinos, the effective number of neutrinos,
the dark energy equation of state, the curvature of the universe,
cosmic strings and isocurvature modes. They have found that the
observations prefer these parameters to be consistent with the
standard values except for one parameter: the effective number of
neutrinos, $N_\nu$. Remarkably, their 2$\sigma$ limit is $N_\nu =
5.3^{+2.1}_{-1.7}$, not allowing the standard value of $N_\nu = 3.0$
at 2.4$\sigma$~\footnote{Ref.~\cite{Mangano:2006ur} recently reexamined this
issue using almost the same data set and found $N_\nu =
4.6^{+1.6}_{-1.5}$ at 95\% C.L. Although the significance is lower
than the one in Ref.~\cite{Seljak:2006bg}, tension with the
standard value remains. Also, there is an independent analysis by
Ref.~\cite{Cirelli:2006kt} with a similar data set including
Lyman-$\alpha$ which gives $N_\nu = 5 \pm 1$, the 2$\sigma$ preference
for $N_\nu > 3$.}.

This preference of a non-standard value of $N_\nu$ by the combined data
of the WMAP3 and Lyman-$\alpha$ forest is closely connected to the
discrepancy of $\sigma_8$ between these data sets.
As shown in Ref.~\cite{Seljak:2006bg}, the larger
value of $N_\nu$ enhances the best fit value of the small scale
amplitude~\footnote{In detail; they report this result in the amplitude at a
smaller scale than 8$h^{-1}$\,Mpc, but similar correlation is expected
between $N_\nu$ and $\sigma_8$.}. This enables the high $\sigma_8$
value inferred from the Lyman-$\alpha$ to reconcile with the WMAP3
data, which prefers the low $\sigma_8$ when the standard $N_\nu = 3$
is assumed. Although there is no detailed statistical analysis of
combined data sets of the WMAP3 and the weak lensing allowing for the
possibility of non-standard $N_\nu$, it is reasonable to expect these
observations too to be reconciled with the WMAP3 data by $N_\nu > 3$.

On the other hand, 
as is very well known, the value of $N_\nu$ greatly affects
big bang nucleosynthesis (BBN), especially the $^4$He abundance,
$Y_p$. The analysis by Ref.~\cite{Mangano:2006ur}, using $Y_p = 0.249
\pm 0.009$ \cite{Olive:2004kq,Cyburt:2004yc}, has yielded $N_\nu =
3.1^{+1.4}_{-1.2}$ (95\% C.L.), in good agreement with the standard
value while still allowing some room for non-standard values. For
example, $N_\nu = 4$, which can better fit the combined data of the
WMAP3 and Lyman-$\alpha$ forest than $N_\nu = 3$, is
acceptable.  However, more recent analyses favor $N_\nu=3$ (see Sec.~\ref{sec:obs} for
more detailed discussion).

Having seen recent observational constraints on $N_\nu$ from the
structure formation and nucleosynthesis, we will now consider how the
value of $N_\nu$ should be in order to satisfy these
constraints. Although the simplest choice would be to have $N_\nu \sim
4$ before and after BBN, concerning the central values, it may be
preferable to have $N_\nu = 3$ during BBN and increase to $N_\nu =$ 4
-- 5 well before the structure formation begins.
 We here note that BBN measures $N_\nu$
around the temperature $T = O({\rm MeV})$ while the structure
formation data tell us $N_\nu$ in a more recent universe, $T \lesssim
100$\,eV, at which the structure formation of the smallest observable
scale (about 1\,Mpc) begins~\footnote{
The present CMB data probe scales larger than $O(10)$\,Mpc or equivalently $T \lesssim 10$\,eV. Meanwhile the CMB alone does not practically constrain $N_\nu$ (the WMAP three-year alone limit is $N_\nu < 42$ at 95\% C.L.~\cite{Ichikawa:2006vm}). 
}.
In terms of the cosmological time, they respectively measure $N_\nu$ around 1\,sec and after $10^8$\,sec. 
Thus, the constraints on $N_\nu$ from BBN and the
structure formation (the CMB, the Lyman-$\alpha$ forest etc.) do not
necessarily coincide at face value in general.

  In this paper, to realize the latter possibility of increasing
$N_\nu$, we investigate models of particles which decay into radiation
between BBN and the structure formation. Candidates would have a
somewhat long lifetime of 1 to $10^8$\,sec after which they decay
``silently", without destroying the light elements, into very light
particles as copious as photons or neutrinos. We show such particles
are found in supersymmetric extensions of theories proposed to solve
the strong CP problem. Namely, we consider the following
possibilities: (a) saxion decay into axions and (b) gravitino decay
into axino and axion.  We also show a candidate present in models
with the right-handed neutrino which are attractive for explaining
neutrino masses. In this case, we consider (c) Dirac right-handed
sneutrino decay into gravitino and right-handed neutrino.  
In the next section we will give a review on the present observational
status on $\sigma_8$ and $N_\nu$ and their possible tensions between
different experiments. 
In Sec.~\ref{sec:models} we give details of the models and parameter space 
where $N_\nu$ is successfully  increased while meeting cosmological constraints. 
Then, Sec.~\ref{sec:conclusion} is devoted to our conclusions and discussion.

\section{Observational tensions in $\sigma_8$ and $N_\nu$}
\label{sec:obs}
 In this section we give a brief review on the several different observations
 and  analyses of $\sigma_8$ and $N_\nu$ and their implications.
 The recent analyses of Lyman-$\alpha$ forest combined with the WMAP
three-year data by two independent groups are in Refs.~\cite{Viel:2006yh}
and \cite{Seljak:2006bg}. The earlier studies by the same groups with
the WMAP first year data are in Refs.~\cite{Seljak:2004xh} and
\cite{Viel:2005ha}. Those analyses used basically the same
Lyman-$\alpha$ forest data sets. Their results seem to consistently
show that $\sigma_8$ derived from the Lyman-$\alpha$ forest data
prefers the higher value of the WMAP first year result rather than the
three-year value, although two groups conclude that there are no
statistically compelling evidence for inconsistency between WMAP3 and
Lyman-$\alpha$ data. In Ref.~\cite{Seljak:2006bg}, around two sigma
discrepancy in the power-law $\Lambda$CDM model was reported but they
concluded that the difference would be explained by a statistical
fluctuation or unknown systematic errors. The analysis of
Ref.~\cite{Viel:2006yh} found weaker significance for the discrepancy
and concluded that the Lyman-$\alpha$ forest data can be in reasonable
agreement with WMAP3. However, it is apparent from the figures of
likelihood contours in Refs.~\cite{Viel:2006yh,Seljak:2006bg} that the
measurements of the small scale fluctuation amplitude by the WMAP3
alone and the Lyman-$\alpha$ forest alone are not fully consistent.

Accumulating data for weak lensing, another efficient probe of
$\sigma_8$, shows a similar tendency. It was first noted in the WMAP3
paper of Ref.~\cite{Spergel:2006hy} that the ground-based weak lensing
measurement by the wide synoptic survey of the Canada-France-Hawaii
Telescope \cite{Hoekstra:2005cs} favors higher values of $\sigma_8
\approx$ 0.8 -- 1.0. In Ref.~\cite{Spergel:2006hy}, the likelihood
contours on the $\Omega_m$-$\sigma_8$ plane are drawn for the WMAP3
alone and the weak lensing alone but their overlapping region is small
showing some degree of inconsistency. Moreover, a higher $\sigma_8$
value is also preferred by a very recently released result of the
space-based measurement by the COSMOS survey of the Hubble Space
Telescope \cite{Massey:2007gh}. The agreement between the largest
surveys from ground and space is remarkable and adds to the
reliability of the weak lensing result of the high $\sigma_8$.

 Regarding the fact that the numerous non-standard parameters other
than $N_\nu$ studied in Ref.~\cite{Seljak:2006bg} cannot solve the
discrepancy, we consider that the universe with $N_\nu > 3$ is a strong
candidate for explaining both the WMAP3 and the observations which indicate high $\sigma_8$.

Now let us look at the constraints on $N_\nu$ from BBN.
The value of $N_\nu$ greatly affects BBN, 
and in particular, the $^4$He abundance $Y_p$ is quite sensitive to it.
 The analysis by Ref.~\cite{Mangano:2006ur}, using $Y_p = 0.249
\pm 0.009$ \cite{Olive:2004kq,Cyburt:2004yc}, has yielded $N_\nu =
3.1^{+1.4}_{-1.2}$ (95\% C.L.), in good agreement with the standard
value while still allowing some room for non-standard values.
However, there are more recent analyses of $Y_p$ by
several groups who give more stringent error bars: $Y_p=0.250 \pm
0.004$ \cite{Fukugita:2006xy}, $Y_p=0.2427 \pm 0.0028$
\cite{Peimbert:2007vm} and $Y_p=0.2516 \pm 0.0011$
\cite{Izotov:2007ed}. We derive the constraints on $N_\nu$ from them
using the fitting formula in Ref.~\cite{Serpico:2004gx} and the
observed deuterium abundance ${\rm D/H} = (2.82 \pm 0.27) \times
10^{-5}$ \cite{O'Meara:2006mj} on the $\eta$-$N_\nu$ plane. The 95\%
C.L. limits are respectively $N_\nu = 3.20^{+0.76}_{-0.68}$,
$N_\nu=3.01^{+0.52}_{-0.48}$ and $N_\nu=3.32^{+0.23}_{-0.24}$ (their
own analysis in Ref.~\cite{Izotov:2007ed} has yielded $N_\nu = 3.28
\pm 0.16$ (2$\sigma$), using $^7$Li data in addition; this is
consistent with our calculation). Although it is beyond the scope of
this paper to discuss whether their error bars are underestimated or
not, we may conclude that three recent analyses of $^4$He do not favor
$N_\nu > 4$. 

Thus we are led to consider a cosmological scenario
that the effective number of neutrinos $N_\nu$ is increased from the standard
value to $N_\nu > 4$ during the time between BBN and the structure 
formation~\footnote{ Strictly
speaking, the $Y_p$ analysis by Ref.~\cite{Izotov:2007ed} implies
$N_\nu > 3$ at 2$\sigma$. However, since our models that we will present
below are not affected by the value of $N_\nu$ at BBN, we assume
$N_\nu = 3$ at BBN for simplicity. }. In the rest of the paper, we will
focus on several possible scenarios based on particle physics and discuss
each model in detail.

\section{Models}
\label{sec:models}

In this section we provide several models in which the effective
number of neutrinos is increased from the standard value, $N_\nu = 3$,
after BBN but before the structure formation starts.  To this end, we
introduce a long-lived particle $X$ with a lifetime $\tau_X$ in the
range of $\tau_X = O(1 $--$10^8)\rm{\,sec}$. The lower bound on
$\tau_X$ comes from the requirement that the additional radiation
energy from the decay of $X$ should not change the expansion rate
before the neutrino freeze-out.  This is because we do not want to
change the standard BBN results, especially the ${}^4$He abundance.
The upper bound corresponds to the cosmic time when the comoving scale
of about $1$\,Mpc enters the horizon.

Before going to the discussion of each model, it will be useful to
express the increase of $N_\nu$ in terms of the abundance and the
lifetime of $X$.  Let us assume that the decay of $X$ produces very
weakly interacting relativistic particles collectively denoted by $R$,
which carry a fraction $f_R$ of the energy originally stored in $X$.
The $R$ particles increase the extra effective number of neutrinos by
$\Delta N_\nu$:
\begin{equation}
	\Delta N_\nu \;\simeq\;3\,f_R \left.\left (
	\frac{\rho_X}{\rho_\nu} \right )\right|_{T=T_d} ,
\label{DeltaN}
\end{equation}
where $\rho_X$ and $\rho_\nu$ denote the energy densities of $X$ and
the three species of the neutrinos, respectively.  We define $T_d$ as
the temperature of photons when the decay rate $\Gamma_X$ becomes
equal to $3H$ ($H$ is the Hubble parameter):
\beq
\Gamma_X \;=\; 3H = \left(\frac{\pi^2 g_*}{10}\right)^{\frac{1}{2}} 
						\frac{T_d^2}{M_P},
\label{GammaX}
\eeq
where $g_* \simeq 3.36$ counts the relativistic degrees of freedom,
and $M_P = 2.4 \times 10^{18}{\rm\,GeV}$ is the reduced Planck mass.
The lifetime $\tau_X$ is related to the decay rate as $\tau_X =
1/\Gamma_X$.  To be precise, speaking, the energy density of the $R$
particles also contributes to the right-hand side of Eq.~(\ref{GammaX}).
Nevertheless we neglect it here, because it is sub-dominant as long as
$\Delta N_\nu \lesssim 1$.

The standard value of the neutrino abundance is
\beq
\left. \frac{\rho_\nu}{s}\right|_{T=T_d} \;\simeq\; 0.26\, T_d,
\label{eq:rho_nu}
\eeq
where $s$ is the entropy density, and it should be noted here that
$T_d$ denotes the temperature of photons, not neutrinos.  Substituting
Eq.~(\ref{eq:rho_nu}) into Eq.~(\ref{DeltaN}), we obtain
\bea
\label{eq:relation}
\Delta N_\nu &\simeq& 1.1\, f_R \lrfp{T_d}{\rm\, keV}{-1} \lrf{\rho_X/s}{10^{-7} {\rm\,GeV}},
\eea
or equivalently,
\bea
\label{eq:relation2}
\Delta N_\nu     &\simeq& 1.2\,f_R \lrfp{\tau_X}{10^6{\rm\, sec}}{\frac{1}{2}}
		         	\lrf{\rho_X/s}{10^{-7}{\rm\,GeV}}.
\eea

It is clear from Eq.~(\ref{eq:relation2}) that, to increase $N_\nu$ by
order unity, $X$ must be produced with a sufficiently large abundance
and its lifetime should be long
enough.  In order not to disturb the light element abundances, the
decay into the standard-model particles (especially into the hadrons)
must be sub-dominant or even forbidden, and the decay products must be
``dark", i.e., their interaction with the visible particles should be
very weak. In other words, any massive particles that decay into very
weakly interacting and relativistic particles can explain the increase
of $N_\nu$ if and only if they have right abundance and lifetime
given by Eq.~(\ref{eq:relation2}).

In the following, we show three cosmologically consistent scenarios in
which $N_\nu$ increases between BBN and structure formation by order
unity. We consider (a) saxion decay into two axions; (b) gravitino
decay into axino and axion; (c) Dirac right-handed sneutrino decay
into gravitino and right-handed neutrino.  We investigate each model
in detail below.

\subsection{Saxion decay into axions}

One of the most promising solutions to the strong CP problem in
quantum chromodynamics (QCD) is the Peccei-Quinn (PQ)
mechanism~\cite{Peccei:1977hh}, which involves a pseudo-Nambu-Goldstone boson $a$, the axion, associated with the spontaneous
PQ symmetry breaking (for a review, see Ref.~\cite{Kim:1986ax}).  In a
supersymmetric theory, the axion forms a supermultiplet, including a
fermionic superpartner $\tilde{a}$, the axino, and a scalar partner
$s$, the saxion.  In general, the saxion acquires a mass of order
$m_{3/2}$ in the presence of the supersymmetry (SUSY)
breaking~\cite{Rajagopal:1990yx,Goto:1991gq}. (Here $m_{3/2}$ is the
gravitino mass.)  In a class of models, the saxion mainly decays into
a pair of the axions (i.e., $f_R \simeq 1$), and these axions
contribute to the extra effective number of neutrinos without
disturbing the BBN results~\footnote{ Note that it is model dependent
whether the saxion dominantly decays into the axions.  For instance,
in a model where two PQ scalar fields $\Phi_+$ and $\Phi_-$,
respectively charged under the PQ symmetry $+1$ and $-1$, acquire VEVs
as $\la \Phi_+ \ra \simeq \la \Phi_- \ra \sim F_a$, the saxion decay
into axions is suppressed.}.  However, since the saxion may also decay
into two photons, we need to examine whether the photons produced do
not spoil the standard cosmology.

We consider a class of models in which the PQ symmetry is spontaneously
broken by a single PQ scalar field $\Phi$, whose vacuum expectation
value (VEV) sets the scale of the PQ symmetry breaking scale $F_a =
\la \Phi \ra$. Here we have assumed that the VEV of $\Phi$ is real and
positive without a loss of generality.  The PQ scale $F_a$ is severely
constrained from astrophysical and cosmological considerations as
$10^{10}{\rm\,GeV} \lesssim F_a \lesssim \theta ^{-1}
10^{12}{\rm\,GeV}$, where $\theta$ is an initial misalignment angle of
the axion.  It is firmly bounded from below by supernova
cooling~\cite{Raffelt:1996wa,Kolb:1990vq}, while the upper bound comes
from the axion-overclosure limit, which can be relaxed to some extent
depending on the cosmological
scenarios~\cite{Preskill:1982cy,Steinhardt:1983ia,Turner:1985si,Kawasaki:1995vt}.

Let us express $\Phi$ in terms of the saxion $s$ and the axion $a$ as
\beq
\Phi \;=\; \frac{s}{\sqrt{2}} \exp\left[i \frac{a}{\la s \ra} \right].
\eeq
Expanding the saxion around its VEV as $s = \sqrt{2} F_a + \hat{s}$, 
we obtain
\bea \Phi &= & \left(F_a+ \frac{\hat{s}}{\sqrt{2}} \right) \exp\left[i
\frac{a}{ \sqrt{2} F_a} \right],\\ \del_\mu \Phi^\dag \del^\mu \Phi
&=& \frac{1}{2} \del_\mu \hat{s} \del^\mu \hat{s} + \frac{1}{2}
\del_\mu a \del^\mu a + \frac{\hat{s}}{\sqrt{2} F_a} \del_\mu a
\del^\mu a + \cdots, \eea
where the third term induces the saxion decay into axions.  The
decay rate is given by
\begin{equation}
	\Gamma (s \to 2a) \;\simeq\; \frac{1}{64\pi}\frac{m_s^3}{F_a^2},
\end{equation}
where $m_s$ is the saxion mass. The lifetime of the saxion then is given as
\begin{equation}
	\tau_s \simeq 1.3 \times 10^5 {\rm\,sec} \left (
	\frac{m_s}{{\rm\,100\,MeV}} \right )^{-3} \left (
	\frac{F_a}{10^{12}{\rm\,GeV}} \right )^{2}.
\label{saxion-tau}	
\end{equation}

Since the axion superfield $\Phi$ must not have a SUSY mass, the
saxion is a flat direction and acquires only a SUSY breaking mass of
the order of the gravitino mass. Therefore, in the early universe, the
initial position of the saxion, $s_i \equiv \sqrt{2}|\Phi_i|$,
naturally deviates from that in the vacuum $\la s \ra = \sqrt{2} F_a$.
When the Hubble parameter becomes comparable to the saxion mass $m_s$,
the saxion starts to oscillate around the potential minimum with an
initial amplitude, $\delta s_i \simeq |s_i - \sqrt{2} F_a|$.  There is
{\it a priori} no way to determine the initial displacement of the saxion,
$\delta s_i$, but it is expected to be in the range between $F_a$ and
$M_P$.

The saxion abundance depends on the thermal history of the universe,
e.g., whether the reheating is completed before or after the saxion
starts to oscillate~\cite{Asaka:1998xa}.  First let us assume that the
saxion starts to oscillate after the reheating.  This is the case if
the reheating temperature $T_R$ satisfies
\begin{equation}
	T_R \;\gtrsim\; 2.2 \times 10^8 {\rm\,GeV}
	\left ( \frac{m_s}{100 {\rm\,MeV}} \right )^{1/2},
	    \label{case(A)condition}
\end{equation}
where we have used the relativistic degrees of freedom in
MSSM, $g_*=228.75$.
The saxion-to-entropy ratio is then given by
\bea
	\frac{\rho_s}{s}&=&\frac{m_s^2 (\delta s_i)^2/2}{3 H_{\rm osc}^2 M_P^2}\
						 \frac{3 T_{\rm osc}}{4} \non\\
	 &\simeq& 4.7 \times 10^{-6} {\rm\,GeV}
	\left ( \frac{m_s}{100{\rm\,MeV}} \right )^{1/2}
	\left ( \frac{F_a}{10^{12}{\rm\,GeV}} \right )^{2} 
         \left ( \frac{\delta s_i}{F_a} \right )^2,
	  \label{rho_s/s:case(A)}
\eea
where the subscript ``osc" denotes that the variables should be
evaluated when the saxion starts to oscillate, i.e., $H \simeq m_s$.
The saxion decays into axions, increasing the effective number of
neutrinos $\Delta N_\nu$ as
\begin{equation}
\begin{split}
	\Delta N_\nu \simeq 2.0 \times 10 
		\left ( \frac{m_s}{100{\rm\,MeV}} \right )^{-1} 
	\left ( \frac{F_a}{10^{12}{\rm\,GeV}} \right )^{3}	
	 \left ( \frac{\delta s_i}{F_a} \right )^2,
 \label{DeltaN_nu:case(A)}
\end{split}
\end{equation}
where we have substituted Eqs.~(\ref{saxion-tau}) and
(\ref{rho_s/s:case(A)}) into Eq.~(\ref{eq:relation2}), and used $f_R
\simeq 1$.
However, as we will see below, it is difficult to reconcile
the constraint on $T_R$, Eq.~(\ref{case(A)condition}), with the gravitino problem.

On the other hand, if the reheating is completed after the oscillation of
the saxion commences, the saxion-to-entropy ratio is given by
\bea
	\frac{\rho_s}{s}&=&\frac{m_s^2 (\delta s_i)^2/2}{3m_s^2 M_P^2}\
						 \frac{3 T_R}{4} \\
	 &\simeq& 2.2 \times 10^{-8}{\rm\,GeV}
	\left ( \frac{T_R}{10^6{\rm\,GeV}} \right )
	\left ( \frac{F_a}{10^{12}{\rm\,GeV}} \right )^{2}
	\left ( \frac{\delta s_i}{F_a} \right )^2.
         \label{rho_s/s:case(B)}
\eea
The increase in the effective number of neutrinos is expressed as
\begin{equation}
\begin{split}
	\Delta N_\nu \simeq 3.0 
	\left ( \frac{m_s}{10{\rm\,MeV}} \right )^{-3/2} 
	\left ( \frac{F_a}{10^{12}{\rm\,GeV}} \right )^{3}
	 \left ( \frac{\delta s_i}{F_a} \right )^2
	\left ( \frac{T_R}{10^6{\rm\,GeV}} \right ).  
	\label{DeltaN_nu:case(B)}
\end{split}
\end{equation}
Thus it is possible to increase $N_\nu$ by order unity in this scenario.

Now let us consider the saxion decay into two photons.  The decay
occurs in the DFSZ axion model \cite{Zhitnitsky:1980tq}, as well as in
the KSVZ (or hadronic) axion model \cite{Kim:1979if} if the heavy
quarks have $U(1)_{\rm em}$ charges.  To be concrete, let us consider
a hadronic axion model by introducing the coupling of $\Phi$ with the
heavy quarks $Q$ and $\bar Q$ as
\beq
W \;=\; k \Phi Q \bar Q,
\label{ksvz}
\eeq
where $k$ is a coupling constant~\footnote{
We assume that the PQ symmetry is broken due to the VEV of $\Phi$ during inflation. Then the PQ quarks $Q$ and $\bar{Q}$ are not thermalized after inflation and they do not affect the timing when the saxion starts oscillating.
}.
We assign the PQ charges as, e.g.,
$\Phi(+1)$, $Q(-1/2)$, and $\bar{Q}(-1/2)$.  Assuming that $Q$ and
$\bar Q$ furnish ${\bf 5}$ and ${\bar {\bf 5}}$ representations of the
$SU(5)$ GUT group, $\Phi$ couples to the standard-model gauge
multiplets as
\beq -{\cal L}_{\rm int} \;=\; \int d^2 \theta \left(
\frac{\alpha_i}{8 \pi} \right) \frac{\Phi}{F_a} W^{(i)}_\alpha W^{(i)
  \alpha} + {\rm h.c.}, \eeq
where $\alpha_i = g_i^2/4\pi$ are the gauge coupling constants of the
standard model, and $W^{(i)}_\alpha$ are chiral superfields for the
gauge multiplets.  Thus the saxion decays into two photons with the
rate,
\begin{equation}
	\Gamma (s \to 2\gamma) \;\simeq\;
			 \frac{\kappa^2 \alpha_{\mathrm{em}}^2}{512\pi ^3}\frac{m_s^3}{F_a^2},
\label{2gamma}			 
\end{equation}
where $\kappa = (3/5)\cos ^2\theta_W \simeq 0.5$ ($\theta_W$ is the
weak mixing angle) and we can see that the branching ratio of two-photon decay is $B_\gamma \simeq 1.7\times 10^{-7}$.
The injected photons may destroy the light elements and change the result of BBN
for $ m_s \gtrsim 40$\,MeV, while for $ m_s \lesssim 40$\,MeV, the
injected photons can never dissociate $^4$He nuclei~\cite{Malcom:1973}.
To avoid changing BBN, the following bounds must be
satisfied~\cite{Kawasaki:1994af,Kawasaki:2004yh}:
\begin{equation}
	B_\gamma \left (\frac{\rho_s}{s} \right ) \; \lesssim \left\{
\bear{ll}
10^{-14} {\rm\, GeV} &{\rm~~for~~} 10^{7}{\rm\, sec} \lesssim \tau_s \lesssim 10^{12}{\rm\, sec} \\
10^{-6}\, \textrm{--}\, 10^{-14}{\rm\, GeV}&{\rm~~for~~} 10^4{\rm\, sec} \lesssim \tau_s \lesssim 10^7{\rm\, sec}
\eear  	
\right. ,
\end{equation}
and the constraints from BBN are very weak for $\tau_s < 10^4$\,sec.
If the saxion mass exceeds about 1\,GeV, the saxion decays into gluons
with the rate
\beq
\Gamma (s \to 2g) \;\simeq\;
			\frac{\alpha_s^2}{64 \pi^3} \frac{m_s^3}{F_a^2},
\eeq
which is much larger than that of Eq.~(\ref{2gamma}).  The hadronic branching
ratio is $B_h = \alpha_s^2/\pi^2 \simeq 1.4 \times 10^{-3}$.  The
bound on the saxion abundance in this case is
\begin{equation}
	B_h \left (\frac{\rho_s}{s} \right ) \;\lesssim \left\{
\bear{cc}
10^{-13}\, \textrm{--}\, 10^{-14} {\rm\, GeV}&{\rm~~~~for~~}10^{4}{\rm\, sec} \lesssim \tau_s \lesssim 
10^{12}{\rm\, sec} \\
10^{-9}\,\, \textrm{--}\,\, 10^{-13} {\rm\, GeV}&{\rm for~~}1{\rm\, sec} \lesssim \tau_s \lesssim 10^{4}{\rm\, sec}
\eear	  	
\right..
\end{equation}
Thus, if the saxion decays into gluons, the BBN constraints on
$\rho_s/s$ become much severer.  In particular, for $m_s > 1$\,GeV the
saxion decay cannot realize $\Delta N_\nu = 1$ due to these
constraints.
Even in the case $m_s \lesssim 40$ MeV,
the energy injection is constrained from the CMB.
If the injected photons cannot reach chemical or kinetic equilibrium 
due to the small rate of interactions with background plasma,
it leads to the distortion of the CMB black body spectrum
which is constrained from observations \cite{Fixsen:1996nj}.
Hence, the constraint comes from the CMB in the region $ m_s \lesssim 40$\,MeV, 
although this does not give a severe constraint.

Let us here comment on the thermal production of the gravitino.  From
Eqs.~(\ref{DeltaN_nu:case(A)}) and (\ref{DeltaN_nu:case(B)}), we can
see that the light saxion mass and/or relatively high reheating
temperature are required to obtain $\Delta N_\nu \sim 1$ as long as we
stick to $\delta s_i \sim F_a$.  Since the saxion mass is considered
to be of the order of the gravitino mass, the gravitino mass as well
must be as light as $O(100)$\,MeV, and such a light gravitino is
realized in gauge-mediated SUSY breaking models~\cite{Giudice:1998bp}.
Let us assume that the gravitino is the lightest supersymmetric
particle (LSP).  If the reheating temperature is too high, the
gravitino may overclose the universe~\cite{Moroi:1993mb}.  The
abundance of the thermally produced gravitino is~\cite{Bolz:2000fu}
(see also~\cite{Kawasaki:2004yh})
\beq
Y_{3/2} \;\simeq\;1.9 \times 10^{-16}
\left ( 1+ \frac{m_{\tilde g}^2}{3m_{3/2}^2} \right )\left ( \frac{T_R}{10^{6}{\rm\,GeV}} \right ),
\label{eq:y32}
\eeq
where we have omitted the logarithmic dependence on $T_R$,
and $m_{\tilde g}$ is the gluino mass evaluated at $T=T_R$.
For $m_{3/2} \ll m_{\tilde g}$, 
the gravitino abundance is given as
\begin{equation}
\begin{split}
	\Omega_{3/2}^\mathrm{TP}h^2 
	\simeq 7.0 \times 10^{-3} \left ( \frac{m_{3/2}}{100{\rm\,MeV}} \right )^{-1}
	\left ( \frac{m_{\tilde g}}{200 {\rm\,GeV}} \right )^{2} 
	 \left ( \frac{T_R}{10^6{\rm\,GeV}} \right ).  \label{TPgravitino}
\end{split}
\end{equation}
This should be smaller than the present upper bound on the current
dark matter density, $\Omega_{\rm DM} h^2 \lesssim 0.12$ at 95\%
C.L.~\cite{Spergel:2006hy}. Therefore, the thermal gravitino
production sets the lower bound on the gravitino mass for a fixed
reheating temperature.  
Due to this bound, there is no allowed region in the case that the reheating is completed before 
the start of saxion oscillations.
It should be noted that this constraint cannot
be alleviated even in the case of the axino
LSP~\cite{Asaka:2000ew}. This is because, although the gravitino
eventually decays into the axion and the axino, the decay is too late
for such a light gravitino mass.  In the next subsection, we consider
a much heavier gravitino mass, focusing on the possibility that the axion
and the axino produced from the decay of gravitino may explain $\Delta
N_\nu \sim 1$.

The axinos, in addition to the gravitinos, are also produced by thermal scattering, and we should check whether the axino is overproduced. Here, since we assume that the axino mass is of the order of the gravitino mass, we do not have to care whether the axino is the LSP or not.
The abundance of the thermally produced axinos is calculated
as~\cite{Covi:2001nw,Brandenburg:2004du},
\begin{equation}
\begin{split}
	Y_{\tilde a}^{\mathrm{TP}}
	\simeq 2.0 \times 10^{-8}
    \left( \frac{\alpha_s (T_R)^3 \ln [0.098 / \alpha_s (T_R)]}{1.1 \times 10^{-4}} \right)   
	\left ( \frac{F_a}{10^{12}{\rm\,GeV}} \right )^{-2} 
	\left ( \frac{T_R}{10^{6}{\rm\,GeV}} \right ) .
\label{TPaxino}
\end{split}
\end{equation}
or equivalently,
\begin{equation}
\begin{split}
	\Omega_{\tilde a}^{\mathrm{TP}}h^2
	\simeq 5.5 \times 10^{-2}
	\left( \frac{\alpha_s (T_R)^3 \ln [0.098 / \alpha_s (T_R)]}{1.1 \times 10^{-4}} \right)   
	\left ( \frac{m_{\tilde a}}{10 {\rm\, MeV}} \right )
	\left ( \frac{F_a}{10^{12}{\rm\,GeV}} \right )^{-2} 
	\left ( \frac{T_R}{10^{6}{\rm\,GeV}} \right ) 
\label{TPaxino_omega}
\end{split}
\end{equation}
where $m_{\tilde a}$ denotes the mass of the axino.
Therefore, if the axino mass is too large and/or the reheating temperature is too high,
the thermally produced axino may overclose the universe.
Using Eqs.~(\ref{DeltaN_nu:case(B)}) and (\ref{TPaxino_omega}),
we derive
\begin{equation}
	\Delta N_\nu \simeq 1.2 
	\left ( \frac{\Omega_{\tilde a}^{\mathrm{TP}} h^2}{0.1} \right )^{-3/2}
	\left ( \frac{m_{\tilde a}}{m_s} \right )^{3/2}
	\left ( \frac{T_R}{10^6\rm\,GeV} \right )^{5/2}
	\left ( \frac{\delta s_i}{F_a} \right )^2.
	\label{eq:axino_overclosure}
\end{equation}
From this equation, the reheating temperature should be less than $10^6$~GeV
as long as $\Delta N_\nu \lesssim 1$ and $\delta s_i \gtrsim F_a$ are assumed. 
Note also that the axino mass in the model (\ref{ksvz}) is smaller than $m_{3/2}$
 unless the PQ scalar has non-minimal coupling with the SUSY breaking sector. So,
the upper bound on $T_R$ may be relaxed if $m_{\tilde a} \ll m_{3/2} \sim m_s$.

\begin{figure}[t]
\begin{center}
\includegraphics[width=1.0\linewidth]{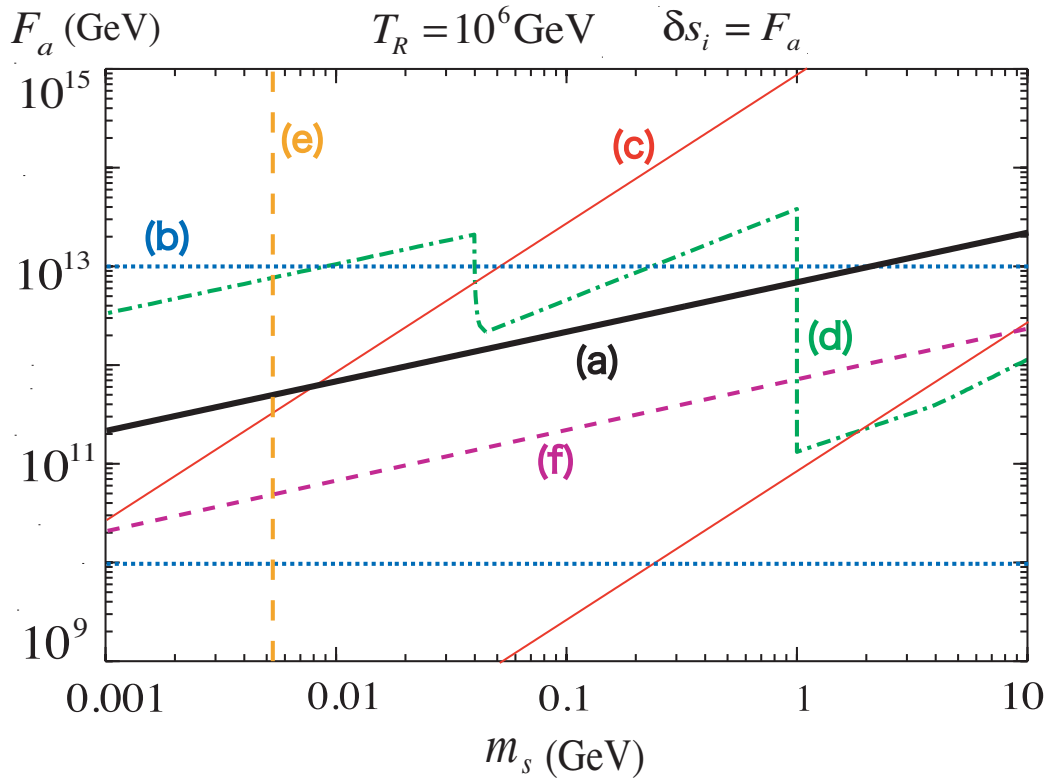} 
\end{center}
\caption{Constraints on the parameter space $m_s$ and $F_a$
in the saxion decay scenario with $T_R = 10^6$\,GeV.
We have chosen $\delta s_i = F_a$. 
The lines labeled (a)--(f) are defined as follows.
(a) $\Delta N_\nu =1$ on this line.  
(b) Lower and upper bounds on the PQ scale with $\theta \sim 0.1$.
(c) Upper line corresponds to $\tau_s \sim 10^8$\,sec,
and lower line corresponds to $\tau_s \sim 1$\,sec. 
(d) BBN bounds coming from radiative decay for $40$\,MeV $\lesssim m_s \lesssim 1$\,GeV and 
hadronic decay for $m_s \gtrsim 1$\,GeV.
For $m_s \lesssim 40$\,MeV, the bound comes from the CMB.
(e) Lower bound on $m_s$ from gravitino thermal production.
(f) Lower bound from axino thermal production for $m_{\tilde a} = 0.01\,m_s$.
For $m_{\tilde a} = m_s$, the constraint coincides with the line (a) accidentally.
}
\label{fig:TR1e6}
\end{figure}

In Fig.~\ref{fig:TR1e6}, we summarize all the constraints discussed
above.  Here we have taken $\delta s_i = F_a$ and $T_R = 10^6$\,GeV as
reference values.  The thick solid black line labeled (a) shows
$\Delta N_\nu = 1$.  Note that the region above this line corresponds
to $\Delta N_\nu \gtrsim 1$.  The dotted blue lines (b) denote the
astrophysical and cosmological constraints on the PQ scale, and we
have set $\theta = 0.1$.  In order to satisfy 1\,sec $\lesssim \tau_s
\lesssim 10^8$\,sec, the combination of parameters ($m_s, F_a$) must
lie in the region between two thin solid red lines (c).  The constraints from
BBN and CMB provide an upper bound on $F_a$ for a fixed $m_s$, as
represented by the dot-dashed green line (d).  The thermally produced
gravitinos exceed the current observed dark matter abundance if $m_s
(\sim m_{3/2}) $ is smaller than the value indicated by the vertical
long-dashed yellow line (e). For $m_{\tilde a} = 0.01\,m_s $, the abundance of the thermally 
produced axinos exceeds the current observed dark matter abundance 
below the long-dashed purple line (f).
For $m_{\tilde a} = m_s$, the constraint from the thermally produced axinos
coincides with the line (a) accidentally, so it is not explicitly drawn.
The region below (a) is excluded  if  $m_{\tilde a} \simeq m_s$.
We have found regions for $\Delta N_\nu \sim 1$
consistent with all the constraints.  For $\delta s_i = F_a$, they
are $1\,$MeV\,$\lesssim m_s \lesssim 1$\,GeV, $F_a\sim 10^{12}$\,GeV
and $10^5$\,GeV\,$\lesssim T_R\lesssim 10^6$\,GeV.  
Since the PQ scale $F_a$ is close to the upper bound coming from the axion-oveclosure
limit, the axion can also play a role of dark matter of the universe.
Moreover, dark matter may be also explained by the thermally produced axinos
(see Eq.~(\ref{eq:axino_overclosure})).

\begin{figure}[t]
\begin{center}
\includegraphics[width=1.0\linewidth]{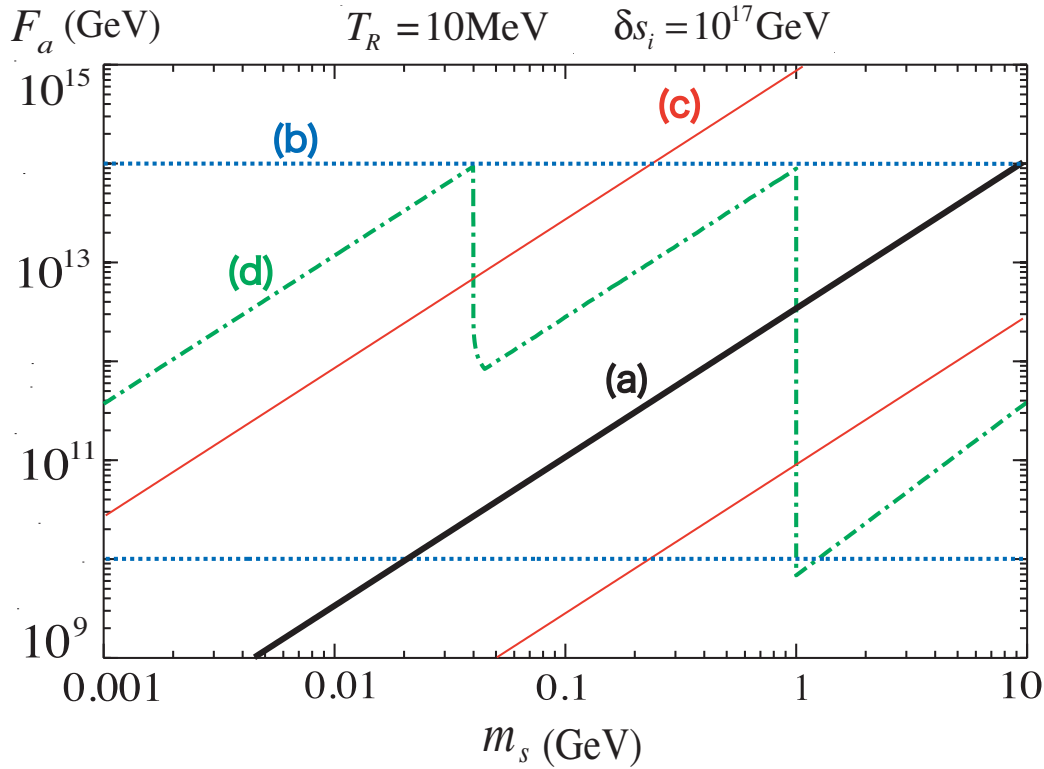} 
\end{center}
\caption{Same as Fig.~\ref{fig:TR1e6}, except for $T_R=10$\,MeV and 
$\delta s_i =10^{17}$\,GeV. }
\label{fig:TR1e-3}
\end{figure}

On the other hand, $\delta s_i$ may be as large as the Planck scale.
We also show the result for $\delta s_i = 10^{17}$\,GeV and $T_R =
10$\,MeV in Fig.~\ref{fig:TR1e-3}, while keeping the other parameters the
same as in Fig.~\ref{fig:TR1e6}.  Note that for such a low reheating
temperature, the axion is diluted and the upper bound on the PQ scale
$F_a$ is relaxed \cite{Kawasaki:1995vt}.  Since $\delta s_i$ is
independent of $F_a$ in this case, the energy density of the saxion is
also independent of $F_a$.  Hence, the BBN and CMB constraints is
given in terms of $\tau _s$ and the dependence on $F_a$ is only
through $\tau_s$, as is seen from Fig.~\ref{fig:TR1e-3}.
For such a low reheating temperature, the axino does not give any meaningful constraint.
For $\delta s_i \sim 10^{16} \textrm{--} 10^{18}$\,GeV, there are allowed
parameter regions, 1\,MeV\,$\lesssim m_s \lesssim 1$\,GeV,
$10^{10}$\,GeV\,$\lesssim F_a \lesssim 10^{14}$\,GeV, and
 a few \,MeV\,$\lesssim T_R \lesssim 100$\,MeV~\footnote{For the precise value of the lower bound on $T_R$,
 see Refs.~\cite{Ichikawa:2006vm,Kawasaki:1999na,Hannestad:2004px,Ichikawa:2005vw}. }.

\subsection{Gravitino decay into axino and axion}  
\label{sec:G->aa}

Next we consider the gravitino decay into the axion and the axino at
late times~\cite{Asaka:2000ew}.  The axino mass is model dependent,
and it can be (much) smaller than the gravitino
mass~\cite{Goto:1991gq}.  Here, from a phenomenological point of view,
we treat the axino mass as a free parameter, but it should be kept in
mind that one may need to contrive a model that realizes a specific
value of the axino mass, especially if it is much smaller than the
gravitino mass.  If the gravitino is the next-to-lightest
supersymmetric particle (NLSP) and the axino is the LSP, the gravitino
decays into the axion and the axino.  Both the axion and the axino
produced by the gravitino decay contribute to the extra effective
number of neutrinos, so $f_R = 1$.

The lifetime of the gravitino is
\begin{equation}
\begin{split}
	\tau(\tilde G \to a +\tilde a) &\; \simeq \;
	\left (\frac{1}{192\pi}\frac{m_{3/2}^3}{M_P^2} \right ) ^{-1}\\
	& \;\simeq\; 8.7\times 10^7 {\rm\,sec}
	\left ( \frac{m_{3/2}}{300{\rm\,GeV}} \right )^{-3}.   \label{tau_Gaa}
\end{split}
\end{equation}
Therefore $m_{3/2}$ must be larger than about 300\,GeV in order to
satisfy the requirement $\tau \lesssim 10^8$\,sec.  The needed
gravitino abundance is given by Eq.~(\ref{eq:relation2}) as
\begin{equation}
	\frac{\rho_{3/2}}{s} \simeq 8.8\times 10^{-9}{\rm\,GeV} \,\,	\Delta N_\nu
	\left ( \frac{m_{3/2}}{300{\rm\,GeV}} \right )^{3/2},
  \label{rho_3/2/s}
\end{equation}
which is given in terms of $Y$ as
\begin{equation}
	Y_{3/2} \simeq 2.9\times 10^{-11} \,\,	\Delta N_\nu
	\left ( \frac{m_{3/2}}{300{\rm\,GeV}} \right )^{1/2}.
   \label{eq:requY_3/2}
\end{equation}
The gravitino may be produced both thermally and non-thermally.  
First we assume that gravitinos are dominantly produced by
particle scatterings in thermal plasma.  From
Eq.~(\ref{eq:y32}), in order to obtain the gravitino abundance
Eq.~(\ref{eq:requY_3/2}), the reheating temperature $T_R$ must be as
high as $O(10^{10})$\,GeV with $m_{\tilde g} \sim O(1)$\,TeV.  For
such a high reheating temperature, however, axinos are also
efficiently produced by thermal scatterings.  Their thermal abundance
is given by Eqs.~(\ref{TPaxino}) or (\ref{TPaxino_omega}).
Thus, for the axino abundance not to exceed the current observed dark matter abundance,  
the axino mass must be smaller than $O(1)$\,keV.

With such a light mass, however, its free
streaming may erase the cosmological structure and conflict with the
observation.
The maximal abundance consistent with the
observational data including Lyman-$\alpha$ forest can be inferred
from the upper bound on the HDM component, or the neutrino masses.
According to Ref.~\cite{Seljak:2006bg}, the 95\% C.L. limit obtained
from the data set including the Lyman-$\alpha$ forest is $\sum m_\nu <
0.17$\,eV which can be converted to $\Omega_\nu h^2 < 1.8 \times
10^{-3}$.  Therefore, it is reasonable to expect that the
contribution to the energy density of the universe from such light axino
must be less than  1\% of the dark matter, in order to be consistent with the
observed Lyman-$\alpha$ forest.
This further constrains the axino mass down to be smaller than $O(10)$\,eV.
Note that, for the axino mass lighter than $O(10)$\,eV,
the axino abundance produced from the gravitino decay is negligibly small.

So far we have assumed that the gravitino with the abundance
Eq.~(\ref{eq:requY_3/2}) is thermally produced. This requires a quite
high reheating temperature, which limits the axino mass to being much
smaller than the gravitino mass.  Since the axino mass is generically
of the order of the gravitino mass, such a hierarchy may pose a
difficulty to build a viable axion model that realizes the axino mass.
If the gravitino is non-thermally produced from, e.g., inflaton
decay~\cite{Kawasaki:2006gs,Endo:2006tf,Endo:2006qk,Endo:2007ih} (or
modulus decay~\cite{Endo:2006zj,Dine:2006ii}), the tension can be
greatly relaxed. The gravitino abundance is then dependent on
the inflaton mass $m_\phi$ and VEV $\la \phi \ra$.  For instance, in a
high scale inflation model, the inflaton decays into the SUSY breaking
sector, producing the gravitino as~\cite{Endo:2007ih}~\footnote{ 
Note that $T_R \sim 10^3$\,GeV is
naturally realized from the spontaneous decay of the inflaton through
the top Yukawa coupling~\cite{Endo:2006qk} for $m_\phi = 10^{12}$\,GeV
and $\la \phi \ra = 10^{15}$\,GeV.  }
\beq
Y_{3/2} \;\simeq\; O( 10^{-11}) \lrfp{T_R}{10^3\,{\rm GeV}}{-1}
		\lrfp{\la \phi \ra}{10^{15}\,{\rm GeV}}{2}
			\lrfp{m_\phi}{10^{12}\,{\rm GeV}}{2},
\label{non-thermal}			
\eeq
where the precise abundance depends
on the details of the SUSY breaking sector. The values adopted for
$m_\phi$ and $\la \phi \ra$ in Eq.~(\ref{non-thermal}) can be realized
in e.g. a hybrid inflation model~\cite{Copeland:1994vg}. For such a
low reheating temperature, the thermal production of the
axino does not set any severe bound on the axino mass. In particular,
note that Eq.~(\ref{TPaxino}) is not applicable for the reheating
temperature smaller than the weak scale. Instead, the
axino produced from the gravitino decay puts an upper bound as
$m_{\tilde a} \lesssim O(100)$\,MeV. This can be derived as follows.
The axino abundance from the gravitino decay is
\beq
\Omega_{\tilde a} h^2 \;\simeq\;8 \times 10^{-4} \Delta N_\nu \lrf{m_{\tilde a}}{100{\rm\, MeV}}
\left ( \frac{m_{3/2}}{300{\rm\, GeV}} \right )^{1/2}.
\label{eq:omega-axino-from-gravitino}
\eeq
Requiring the axino abundance to be
smaller than 1\% of the dark matter,
we obtain
an upper bound on the axino mass as $m_{\tilde{a}} \lesssim O(100)\, {\rm MeV}$.
Although the axino mass cannot be as large as the gravitino mass, the
required hierarchy of the two is rather mild, compared to the
previous case. In a similar fashion, we can show that the
LSPs produced by this non-thermal process cannot be the dominant
component of the dark matter in the models described below.

In the present model, both the axion and axino are produced from the
gravitino decay as relativistic particles.
In contrast to the axion, the axino becomes non-relativistic at some time
depending on its mass.  But it is typically well after the
matter-radiation equality epoch, and the axino abundance amounts to
only a small fraction of the energy density of the universe. Thus both
the axion and the axino contribute to the effective number of the
neutrinos, that is, $f_R = 1$. The same argument is applied to the
following model as well.

So far we have neglected the saxion abundance. Since the saxion mass
is roughly equal to the gravitino mass, the saxion decays much earlier
than BBN begins (see Eq.~(\ref{saxion-tau})). In addition, if $\delta
s_i$ is of $O(F_a)$, since the saxion does not dominate the universe,
our arguments above remain unchanged.

The final comment is that one cannot exchange the roles of the gravitino
and the axino in the above scenario.  Similar arguments show that the
gravitino must be much lighter than the axino. This is because, as long as
we require $\Delta N_\nu \simeq 1$,  the gravitinos produced from the axino 
are so abundant that the small scale fluctuations ($\gtrsim$ a few Mpc) 
would be smoothed out unless the gravitino mass is small enough.  However, 
since the axion multiplet cannot have a SUSY mass, the axino mass 
cannot be much larger than the gravitino mass and the scenario does not seem
to work.

\subsection{Dirac right-handed sneutrino decay into gravitino and right-handed neutrino}

The neutrino oscillation experiments have revealed that  the neutrinos have
finite but small masses.  To explain the tiny neutrino masses one introduces
right-handed neutrinos into the standard model.  
The right-handed neutrinos may be allowed to have
large Majorana masses as large as GUT scale, because they are singlets
with respect to the standard-model gauge group.  However, the Majorana
mass term can be forbidden by some symmetry such as the lepton-number
symmetry. 
Thus, if this is the case, the neutrino mass is given by the Dirac mass
term, and the mass of the right-handed neutrino is very light.  The smallness
of the neutrino mass is explained by the small Yukawa coupling of 
$\sim m_\nu / \langle H_u \rangle \lesssim O(10^{-13})$,
where $m_\nu$ is a neutrino mass and 
$\langle H_u \rangle$ denotes the VEV of the up-type Higgs.
On the other hand, the right-handed sneutrino acquires a mass  from SUSY
breaking effects.  Since the Yukawa coupling is rather small,
the lifetime of right-handed sneutrinos is very long, and their decay
into the right-handed neutrino and the gravitino can increase $ N_\nu $.

First, the lifetime of right-handed sneutrinos is given as
\begin{equation}
\begin{split}
	\tau _{\tilde \nu _R} &\simeq
	\left ( \frac{1}{48\pi} \frac{m_{\tilde \nu _R}^5}{m_{3/2}^2 M_P^2} \right )^{-1}\\
	& \simeq 1.4 \times 10^8 {\rm\,sec}
	\left ( \frac{m_{3/2}}{500{\rm\,keV}} \right )^2
	\left ( \frac{m_{\tilde \nu _R}}{1 {\rm\,GeV}} \right )^{-5},  \label{tau_aGa}
\end{split}
\end{equation}
where the right-handed sneutrino mass $m_{\tilde \nu _R}$ should be less than $\sim 1$\,GeV as
discussed later.  From Eq.~(\ref{eq:relation2}), the abundance of the right-handed sneutrinos
should be
\begin{equation}
\begin{split}
	\frac{\rho_{\tilde \nu _R}}{s} 
	 \simeq 6.8 \times 10^{-9} {\rm\,GeV} ~ \Delta N_\nu
	\left ( \frac{m_{3/2}}{500{\rm\,keV}} \right )^{-1}
	\left ( \frac{m_{\tilde \nu _R}}{1 {\rm\,GeV}} \right )^{5/2}. 
 \label{axino/s}
\end{split}
\end{equation}
Such large abundance of the  sneutrino is
unlikely to be produced by thermal scatterings or decays of other
superparticles due to the smallness of $m_{\tilde \nu
_R}$ and the Yukawa coupling~\cite{Asaka:2005cn}.  But, the sufficient energy density of
right-handed sneutrino can be non-thermally produced in the form of 
coherent oscillations.  The right-handed
sneutrino can develop a large field value during inflation, and after
the inflation ends it begins to oscillate coherently, which can induce
a large abundance of the right-handed sneutrino~\cite{Murayama:1993em,McDonald:2006if}.  
Its energy
density-to-entropy ratio $\rho_{\tilde \nu_R}/s$ is fixed at the end
of the reheating process,
\begin{equation}
\begin{split}
	\frac{\rho_{\tilde \nu_R}}{s} &=
	\frac{m_{\tilde \nu_{R}}^2|\tilde \nu_{Ri}|^2 T_R}{4H_{{\rm osc}}^2M_P^2} \\
	&\simeq 4.3\times 10^{-8}{\rm\,GeV}
	\left ( \frac{ |\tilde \nu_{Ri}| }{10^{14}{\rm\,GeV}} \right )^{2}
	\left ( \frac{T_R}{100 {\rm\,GeV}} \right ),   \label{coherentnu}
\end{split}
\end{equation}
where $\tilde \nu_{Ri}$ denotes the initial amplitude of the
right-handed sneutrino.  Here we used the Hubble parameter $H_{{\rm
osc}}$ at the start of the oscillations is equal to $m_{\tilde
\nu_{R}}$.

As discussed above, the abundance of gravitinos produced by $\tilde
\nu _R$ decay should be subdominant component of the dark matter.  The
abundance is given by
\begin{equation}
\begin{split}
	\Omega_{3/2} h^2
	 \simeq 9.3 \times 10^{-4} 
	\left ( \frac{m_{\tilde \nu _R}}{1{\rm\,GeV}} \right )^{3/2}.
\label{NTgravitino}
\end{split}
\end{equation}
It is interesting that the abundance is independent of $m_{3/2}$.
In order not to significantly affect the observed small scale structure ($\gtrsim$ a few Mpc),
$m_{\tilde \nu _R}$ must be smaller than $ \sim 1$\,GeV.
With the constraint $\tau \lesssim 10^8$\,sec,
$m_{3/2}\lesssim 500$\,keV is also required.
Note that such a hierarchical mass relation may be realized in 
gauge-mediated SUSY breaking models
with some extended gauge interaction
which involves right-handed neutrinos and is broken at an intermediate scale.

Finally, we comment on the gravitino decay into $\tilde \nu_R$ and $\nu_R$.
This case leads to the same result as in the previous section,
after exchanging ($\tilde \nu_R, \nu_R$) with ($\tilde a , a $).
Therefore, for $\Delta N_\nu \sim 1$,
$m_{3/2} \gtrsim 300$\,GeV and $m_{\tilde \nu_R}\lesssim 1$\,MeV are required.
However, this hierarchical mass relation, $m_{3/2} \gg m_{\tilde \nu_R}$
is unlikely in SUSY models.
Hence, this case is not expected to explain increasing $ \Delta N_\nu$.

\section{Conclusions and discussion}
\label{sec:conclusion}

In this paper, we present models of decaying particles for increasing the
effective number of neutrinos $N_\nu$ after BBN but before the
structure formation begins. 
In the model (a) where the saxion decays into two axions, broad regions are allowed.
For instance, $T_R$ can take a value from 10\,MeV up to $10^6$\,GeV,
depending on the initial displacement of the saxion and $m_{\tilde a} / m_s $
(see Figs.~\ref{fig:TR1e6} and \ref{fig:TR1e-3} for details).
In particular, the saxion mass needs to lie in
the range between 1\,MeV and 1\,GeV, which suggests the light gravitino.
In the model (b) where the gravitino decays into the axino and the axion,
we require $m_{3/2} \gtrsim 300$\,GeV together with $m_{\tilde a} \lesssim 10$\,eV
or $m_{\tilde a} \lesssim 100$\,MeV depending on the gravitino production processes.
The former (latter) bound is the case with the thermal (non-thermal) production.
In particular, one needs a hierarchy
between the gravitino mass and the axino mass.
In the model (c) where the Dirac right-handed sneutrino decays into the 
gravitino and the right-handed neutrino,
$m_{\tilde \nu_R} \lesssim 1$\,GeV and $m_{3/2} \lesssim 500$\,keV are required.
This case works only for the non-thermal origin of the right-handed sneutrino in the form of
scalar condensates. 

Such a scenario is motivated because
non-standard values of $N_\nu > 3$ are preferred by the combined data
of the CMB, galaxy clustering and the Lyman-$\alpha$ forest
\cite{Seljak:2006bg,Mangano:2006ur,Cirelli:2006kt} whereas most of the
recent analyses of primordial $^4$He abundance favor standard $N_\nu =
3$ \cite{Olive:2004kq,Fukugita:2006xy,Peimbert:2007vm}. As is
discussed in Ref.~\cite{Seljak:2006bg}, the preference for $N_\nu > 3$
of the Lyman-$\alpha$ combined data stems from the inconsistency in
the estimation of the matter power spectrum amplitude at small scales,
represented by $\sigma_8$, between the WMAP and the Lyman-$\alpha$
forest: the latter yields somewhat higher value of $\sigma_8$. We note
that such higher $\sigma_8$ values are also derived by other probes of
the small scale matter power spectrum by the weak lensing
\cite{Hoekstra:2005cs,Massey:2007gh} and strong lensing
\cite{Li:2006tk}. Thus, we would like to stress that the models
proposed here can not only solve the discrepancy between BBN and the
structure formation (the CMB, the Lyman-$\alpha$ forest and so on) but
also give a possible answer to the inconsistency between the WMAP and
small scale matter power measurements such as the Lyman-$\alpha$
forest and weak lensing.

It should be emphasized that $N_\nu$ is increased by the
``free-streaming" relativistic particles like massless neutrinos in
our models.  Our scenario of increasing $N_\nu$ may recall the readers
to the scenario of ``interacting" neutrinos discussed e.g. in
Refs.~\cite{Chacko:2003dt,Bell:2005dr} whose prediction includes the
increase in $N_\nu$ after BBN by the recoupling. Even though $N_\nu$
changes by the same amount, there is a stark contrast between the
free-streaming particles and interacting ones as regards the effects
on the structure formation. The consequence is that, although $N_\nu$
can be increased in the interacting neutrino scenario, it cannot solve
the problem.  This is explicitly verified in
Ref.~\cite{Cirelli:2006kt}. Their Fig.~5 (a) shows that the
free-streaming particles can better fit the Lyman-$\alpha$ data by
increasing $N_\nu$ from 3 but such is not the case for interacting
particles as shown in Fig.~5 (b).

Finally, since the discrepancy which we have addressed in this paper
is about 2$\sigma$ level, further data and studies are necessary in
order to see whether inconsistency exists in the standard cosmological
model or in the interpretation of one or more observations. Recently,
Ref.~\cite{Ichikawa:2006vm} have obtained the constraint on $N_\nu$
from the WMAP
\cite{Spergel:2006hy,Page:2006hz,Hinshaw:2006ia,Jarosik:2006ib} and
the SDSS luminous red galaxy power spectrum \cite{Tegmark:2006az} to
be $0.9 < N_\nu < 8.2$ (95\% C.L.). This is not in conflict with the
one derived using the Lyman-$\alpha$ and the earlier galaxy power
spectrum as mentioned in the Introduction, $N_\nu = 4.6^{+1.6}_{-1.5}$
\cite{Mangano:2006ur}, but it does not have sufficient sensitivity to
test the need for $N_\nu > 3$. Since we cannot expect the galaxy power
spectrum data to increase significantly in near future, improvement in
the reliability of the Lyman-$\alpha$ forest and weak lensing will be
needed to solve the issue. We believe the ongoing works in the
communities to understand sources of systematic errors will accomplish
this task and, together with the future CMB experiments (the PLANCK
sensitivity for $N_\nu$ is forecasted to be $\sim 0.2$, see
e.g. \cite{Ichikawa:2006dt}), this will tell us whether the scenario of
increasing $N_\nu$ is realized in Nature.

\section*{Acknowledgements}

The works of KI, MK and MS were supported in part by a
Grant-in-Aid of the Ministry of Education,
Culture, Sports, Science, and Technology, Government of Japan
(No. 18840010 for KI, No. 18540254 and 14102004 for MK, and No. 18840011 for MS).
This work was also supported in part by the
JSPS-AF Japan-Finland Bilateral Core Program.



\begin{thebibliography}{}

\bibitem{Spergel:2006hy}
    D.~N.~Spergel {\it et al.},
    arXiv:astro-ph/0603449.

\bibitem{Page:2006hz}
    L.~Page {\it et al.},
    arXiv:astro-ph/0603450.

\bibitem{Hinshaw:2006ia}
    G.~Hinshaw {\it et al.},
    arXiv:astro-ph/0603451.

\bibitem{Jarosik:2006ib}
    N.~Jarosik {\it et al.},
    arXiv:astro-ph/0603452.

\bibitem{Tegmark:2003uf}
  M.~Tegmark {\it et al.}  [SDSS Collaboration],
  Astrophys.\ J.\  {\bf 606}, 702 (2004).

\bibitem{Tegmark:2006az}
  M.~Tegmark {\it et al.},
  Phys.\ Rev.\ D {\bf 74}, 123507 (2006).

\bibitem{McDonald:2004eu}
  P.~McDonald {\it et al.},
  Astrophys.\ J.\ Suppl.\  {\bf 163}, 80 (2006).

\bibitem{McDonald:2004xn}
  P.~McDonald {\it et al.},
  Astrophys.\ J.\  {\bf 635}, 761 (2005).

\bibitem{Viel:2006yh}
  M.~Viel, M.~G.~Haehnelt and A.~Lewis,
  Mon.\ Not.\ Roy.\ Astron.\ Soc.\ Lett.\  {\bf 370}, L51 (2006).

\bibitem{Seljak:2006bg}
  U.~Seljak, A.~Slosar and P.~McDonald,
  JCAP {\bf 0610}, 014 (2006).

\bibitem{Hoekstra:2005cs}
  H.~Hoekstra {\it et al.},
    Astrophys.\ J.\  {\bf 647}, 116 (2006).

\bibitem{Massey:2007gh}
  R.~Massey {\it et al.},
  arXiv:astro-ph/0701480.

\bibitem{Li:2006tk}
  G.~L.~Li, S.~Mao, Y.~P.~Jing, H.~J.~Mo, L.~Gao and W.~P.~Lin,
  Mon.\ Not.\ Roy.\ Astron.\ Soc.\ Lett.\  {\bf 372}, L73 (2006).

\bibitem{Spergel:2003cb}
  D.~N.~Spergel {\it et al.}  [WMAP Collaboration],
  Astrophys.\ J.\ Suppl.\  {\bf 148}, 175 (2003).

\bibitem{Mangano:2006ur}
  G.~Mangano, A.~Melchiorri, O.~Mena, G.~Miele and A.~Slosar,
  JCAP {\bf 0703}, 006 (2007).

\bibitem{Cirelli:2006kt}
  M.~Cirelli and A.~Strumia,
  JCAP {\bf 0612}, 013 (2006).

\bibitem{Olive:2004kq}
  K.~A.~Olive and E.~D.~Skillman,
  Astrophys.\ J.\  {\bf 617}, 29 (2004).

\bibitem{Cyburt:2004yc}
  R.~H.~Cyburt, B.~D.~Fields, K.~A.~Olive and E.~Skillman,
  Astropart.\ Phys.\  {\bf 23}, 313 (2005).

\bibitem{Ichikawa:2006vm}
  K.~Ichikawa, M.~Kawasaki and F.~Takahashi,
  arXiv:astro-ph/0611784.

\bibitem{Seljak:2004xh}
  U.~Seljak {\it et al.}  [SDSS Collaboration],
  Phys.\ Rev.\ D {\bf 71}, 103515 (2005).

\bibitem{Viel:2005ha}
  M.~Viel and M.~G.~Haehnelt,
  Mon.\ Not.\ Roy.\ Astron.\ Soc.\  {\bf 365}, 231 (2006).

\bibitem{Fukugita:2006xy}
  M.~Fukugita and M.~Kawasaki,
  Astrophys.\ J.\  {\bf 646}, 691 (2006).

\bibitem{Peimbert:2007vm}
  M.~Peimbert, V.~Luridiana and A.~Peimbert,
  arXiv:astro-ph/0701580.

\bibitem{Izotov:2007ed}
  Y.~I.~Izotov, T.~X.~Thuan and G.~Stasinska,
  arXiv:astro-ph/0702072.

\bibitem{Serpico:2004gx}
  P.~D.~Serpico, S.~Esposito, F.~Iocco, G.~Mangano, G.~Miele and O.~Pisanti,
  JCAP {\bf 0412}, 010 (2004).

\bibitem{O'Meara:2006mj}
  J.~M.~O'Meara, S.~Burles, J.~X.~Prochaska, G.~E.~Prochter, R.~A.~Bernstein and K.~M.~Burgess,
  Astrophys.\ J.\  {\bf 649}, L61 (2006).



\bibitem{Peccei:1977hh}
  R.~D.~Peccei and H.~R.~Quinn,
  Phys.\ Rev.\ Lett.\  {\bf 38}, 1440 (1977).


\bibitem{Kim:1986ax}
  J.~E.~Kim,
  Phys.\ Rept.\  {\bf 150}, 1 (1987).


\bibitem{Rajagopal:1990yx}
  K.~Rajagopal, M.~S.~Turner and F.~Wilczek,
  Nucl.\ Phys.\ B {\bf 358}, 447 (1991).


\bibitem{Goto:1991gq}
  T.~Goto and M.~Yamaguchi,
  Phys.\ Lett.\ B {\bf 276}, 103 (1992);\\
  E.~J.~Chun, J.~E.~Kim and H.~P.~Nilles,
  Phys.\ Lett.\ B {\bf 287}, 123 (1992);\\
  E.~J.~Chun and A.~Lukas,
  Phys.\ Lett.\ B {\bf 357}, 43 (1995).

\bibitem{Raffelt:1996wa}
  G.~G. Raffelt, {\em Stars as laboratories for fundamental physics: The
    astrophysics of neutrinos, axions, and other weakly interacting  
    particles}.
  \newblock The University of Chicago Press, Chicago \& London, 1996.

\bibitem{Kolb:1990vq}
  E.~W. Kolb and M.~S. Turner, {\em The Early Universe}.
  \newblock Addison-Wesley, Redwood City, USA, 1990.

\bibitem{Preskill:1982cy}
  J.~Preskill, M.~B.~Wise and F.~Wilczek,
  Phys.\ Lett.\ B {\bf 120}, 127 (1983);\\
  L.~F.~Abbott and P.~Sikivie,
  Phys.\ Lett.\ B {\bf 120}, 133 (1983);\\
  M.~Dine and W.~Fischler,
  Phys.\ Lett.\ B {\bf 120}, 137 (1983).

\bibitem{Steinhardt:1983ia}
  P.~J.~Steinhardt and M.~S.~Turner,
  Phys.\ Lett.\ B {\bf 129}, 51 (1983).

\bibitem{Turner:1985si}
  M.~S.~Turner,
  Phys.\ Rev.\ D {\bf 33}, 889 (1986).

\bibitem{Kawasaki:1995vt}
  M.~Kawasaki, T.~Moroi and T.~Yanagida,
  Phys.\ Lett.\ B {\bf 383}, 313 (1996).


\bibitem{Asaka:1998xa}
  T.~Asaka and M.~Yamaguchi,
  Phys.\ Rev.\  D {\bf 59}, 125003 (1999).


\bibitem{Zhitnitsky:1980tq}
  A.~R.~Zhitnitsky,
  Sov.\ J.\ Nucl.\ Phys.\  {\bf 31} (1980) 260;\\
  M.~Dine, W.~Fischler and M.~Srednicki,
  Phys.\ Lett.\ B {\bf 104}, 199 (1981).

\bibitem{Kim:1979if}
  J.~E.~Kim,
  Phys.\ Rev.\ Lett.\  {\bf 43}, 103 (1979);\\
  M.~A.~Shifman, A.~I.~Vainshtein and V.~I.~Zakharov,
  Nucl.\ Phys.\ B {\bf 166}, 493 (1980).


\bibitem{Malcom:1973}
  C.~K.~Malcom, D.~V.~Webb, Y.~M.~Shin and D.~M.~Skopik,
  Phys.\ Lett.\ B {\bf 47}, 433 (1973).


\bibitem{Fixsen:1996nj}
  D.~J.~Fixsen, E.~S.~Cheng, J.~M.~Gales, J.~C.~Mather, R.~A.~Shafer and E.~L.~Wright,
  Astrophys.\ J.\  {\bf 473}, 576 (1996);\\
G.~F.~Smoot and D.~Scott,
arXiv:astro-ph/9711069.

\bibitem{Kawasaki:1994af}
  M.~Kawasaki and T.~Moroi,
  Prog.\ Theor.\ Phys.\  {\bf 93}, 879 (1995);\\
  E.~Holtmann, M.~Kawasaki, K.~Kohri and T.~Moroi,
  Phys.\ Rev.\ D {\bf 60}, 023506 (1999);\\
  M.~Kawasaki, K.~Kohri and T.~Moroi,
  Phys.\ Rev.\ D {\bf 63}, 103502 (2001);\\
  R.~H.~Cyburt, J.~R.~Ellis, B.~D.~Fields and K.~A.~Olive,
  Phys.\ Rev.\ D {\bf 67}, 103521 (2003).

\bibitem{Kawasaki:2004yh}
  M.~Kawasaki, K.~Kohri and T.~Moroi,
  Phys.\ Lett.\ B {\bf 625}, 7 (2005); 
  Phys.\ Rev.\ D {\bf 71}, 083502 (2005).

\bibitem{Giudice:1998bp}
  For a review, see G.~F.~Giudice and R.~Rattazzi,
  Phys.\ Rept.\  {\bf 322}, 419 (1999).

\bibitem{Moroi:1993mb}
  T.~Moroi, H.~Murayama and M.~Yamaguchi,
  Phys.\ Lett.\ B {\bf 303}, 289 (1993).

\bibitem{Bolz:2000fu}
  M.~Bolz, A.~Brandenburg and W.~Buchmuller,
  Nucl.\ Phys.\ B {\bf 606}, 518 (2001);\\
  J.~Pradler and F.~D.~Steffen,
  arXiv:hep-ph/0612291;\\
  V.~S.~Rychkov and A.~Strumia,
  arXiv:hep-ph/0701104.


\bibitem{Asaka:2000ew}
  T.~Asaka and T.~Yanagida,
  Phys.\ Lett.\ B {\bf 494}, 297 (2000).


\bibitem{Covi:2001nw}
  L.~Covi, H.~B.~Kim, J.~E.~Kim and L.~Roszkowski,
  JHEP {\bf 0105}, 033 (2001).

\bibitem{Brandenburg:2004du}
  A.~Brandenburg and F.~D.~Steffen,
  JCAP {\bf 0408}, 008 (2004).



\bibitem{Kawasaki:1999na}
  M.~Kawasaki, K.~Kohri and N.~Sugiyama,
  Phys.\ Rev.\ Lett.\  {\bf 82}, 4168 (1999);\\
  M.~Kawasaki, K.~Kohri and N.~Sugiyama,
  Phys.\ Rev.\  D {\bf 62}, 023506 (2000).


\bibitem{Hannestad:2004px}
  S.~Hannestad,
  Phys.\ Rev.\  D {\bf 70}, 043506 (2004).


\bibitem{Ichikawa:2005vw}
  K.~Ichikawa, M.~Kawasaki and F.~Takahashi,
  Phys.\ Rev.\  D {\bf 72}, 043522 (2005).


\bibitem{Kawasaki:2006gs}
  M.~Kawasaki, F.~Takahashi and T.~T.~Yanagida,
  Phys.\ Lett.\ B {\bf 638}, 8 (2006);
  Phys.\ Rev.\ D {\bf 74}, 043519 (2006).

\bibitem{Endo:2006tf}
  M.~Endo, K.~Hamaguchi and F.~Takahashi,
  Phys.\ Rev.\  D {\bf 74}, 023531 (2006).

\bibitem{Endo:2006qk}
  M.~Endo, M.~Kawasaki, F.~Takahashi and T.~T.~Yanagida,
  Phys.\ Lett.\  B {\bf 642}, 518 (2006).

\bibitem{Endo:2007ih}
  M.~Endo, F.~Takahashi and T.~T.~Yanagida,
  arXiv:hep-ph/0701042.

\bibitem{Endo:2006zj}
  M.~Endo, K.~Hamaguchi and F.~Takahashi,
  Phys.\ Rev.\ Lett.\  {\bf 96}, 211301 (2006);\\
  S.~Nakamura and M.~Yamaguchi,
  Phys.\ Lett.\  B {\bf 638}, 389 (2006).

\bibitem{Dine:2006ii}
  M.~Dine, R.~Kitano, A.~Morisse and Y.~Shirman,
  Phys.\ Rev.\  D {\bf 73}, 123518 (2006).

\bibitem{Copeland:1994vg}
  E.~J.~Copeland, A.~R.~Liddle, D.~H.~Lyth, E.~D.~Stewart and D.~Wands,
  Phys.\ Rev.\ D {\bf 49}, 6410 (1994);\\
  G.~R.~Dvali, Q.~Shafi and R.~K.~Schaefer,
  Phys.\ Rev.\ Lett.\  {\bf 73}, 1886 (1994);\\
  A.~D.~Linde and A.~Riotto,
  Phys.\ Rev.\ D {\bf 56}, 1841 (1997).


\bibitem{Asaka:2005cn}
  T.~Asaka, K.~Ishiwata and T.~Moroi,
  Phys.\ Rev.\ D {\bf 73}, 051301 (2006);
  Phys.\ Rev.\  D {\bf 75}, 065001 (2007).

\bibitem{Murayama:1993em}
  H.~Murayama and T.~Yanagida,
  Phys.\ Lett.\ B {\bf 322}, 349 (1994).

\bibitem{McDonald:2006if}
  J.~McDonald,
  JCAP {\bf 0701}, 001 (2007).

\bibitem{Chacko:2003dt}
  Z.~Chacko, L.~J.~Hall, T.~Okui and S.~J.~Oliver,
  Phys.\ Rev.\  D {\bf 70}, 085008 (2004).

\bibitem{Bell:2005dr}
  N.~F.~Bell, E.~Pierpaoli and K.~Sigurdson,
  Phys.\ Rev.\  D {\bf 73}, 063523 (2006).

\bibitem{Ichikawa:2006dt}
  K.~Ichikawa and T.~Takahashi,
  Phys.\ Rev.\  D {\bf 73}, 063528 (2006).

\end{thebibliography}
\end{document}